\magnification\magstep1 
  \baselineskip = 0.6 true cm
                           
  \def\sa{\vskip 0.30 true cm}
  \def\sb{\vskip 0.60 true cm}

  \pageno=0
         

   \hoffset = 2.5true mm

   \hsize = 16.2 true cm



\def\souligne#1{$\underline{\smash{ \hbox{#1}}}$}

 \rightline{\bf LYCEN 9222}
 \rightline{May 1992}

\vglue 1 true cm
\sa
\sb
\sa

\noindent 
{\bf Symmetry adaptation techniques in $n$-photon absorption spectroscopy}

\sa
\sb

\noindent 
{M.~Daoud and \souligne{M.~Kibler}}

\sa

\noindent Institut de Physique Nucl\'eaire, IN2P3-CNRS et 
Universit\'e Lyon-1, 69622 Villeurbanne Cedex, France

\sa
\sa
\sa
\sa
\sb
\sb
\sb
\sb

\noindent
Communication to the International Conference {\bf Rare 
Earths' 92 in Kyoto}, Kyoto, Japan, 1-5 June 1992 
(contribution to ``The Caro and Judd Symposium~-~Spectroscopy 
of Rare Earths'').  

\sa
\sb

\noindent 
Paper published in {\bf Journal of Alloys and 
Compounds 193 (1993) 219} (Proceedings of the 
International Conference Rare Earths' 92 in Kyoto). All 
correspondence concerning this paper should be addressed to~:
M.~Kibler, Institut de Physique Nucl\'eaire, Universit\'e 
Lyon-1, 43 Bd du 11 Novembre 1918, 
69622 Villeurbanne Cedex, France. Telephone~: (33) 72 
44 82 35. Fax~: (33) 72 44 80 04. (Electronic mail~: 
kibler@frcpn11) 

\sa
\sb

\sa
\sb

    \vfill\eject

\noindent 
{\bf Symmetry adaptation techniques in $n$-photon absorption spectroscopy}

\sa
\sb

\noindent 
{M.~Daoud and {M.~Kibler}}

\sa

\noindent Institut de Physique Nucl\'eaire, IN2P3-CNRS et 
Universit\'e Lyon-1, 69622 Villeurbanne Cedex, France

\sa
\sb

\baselineskip = 0.50 true cm

\noindent
{\bf Summary}

We present some recent progress achieved in the application of 
symmetry adaptation techniques to $n$-photon absorption 
spectroscopy of rare earth ions in finite symmetry. More 
specifically, 
this work is concerned with the determination of the intensity 
of $n$-photon transitions between Stark levels (rather than $J$ 
levels) of an ion in an environment with a given 
symmetry. The role of symmetry is taken into account through 
the use of initial and final state vectors characterized by 
irreducible representations of the (double) group for the 
ion site symmetry. Two distinct situations are considered, viz., the 
case of parity-allowed $n$-photon transitions (as 
e.g.~intra-configurational transitions between Stark levels of the 
$4f^N$ configuration for $n$ even) and the case of 
parity-forbidden $n$-photon transitions (as 
e.g.~inter-configurational transitions from Stark levels of the 
$4f^N$ configuration to Stark levels of the 
$4f^{N-1}5d$ configuration for $n$ even). 

    \baselineskip = 0.820 true cm

\noindent 
{\bf 1. Introduction}

One- and two-photon spectroscopy [1-22] is a useful 
experimental tool for investigating the electronic properties 
of rare earth ions in crystals. The basic corresponding 
theoretical models have been developed by Judd [1] and Ofelt 
[2] for intra-configurational one-photon transitions (see [20] 
for a recent review) and by Axe [3] for intra-configurational 
two-photon transitions. The case of two-photon transitions has 
been the object of further works and extensions~: (i) 
introduction of higher-order mechanisms for 
intra-configurational transitions [6,7,9,11], (ii) development of 
models for inter-configurational transitions [10,12,15,21], and (iii) 
introduction of symmetry considerations from both a qualitative 
[4,5] and a quantitative [13,14,18,19,21,22] viewpoint. 

In this paper, we present a theoretical model for $n$-photon 
dipolar transitions 
which includes, in a symmetry adapted form, the special 
cases $n = 1$ [1,2] and $n=2$ 
[3]. The model is developed for an arbitrary configuration of $N$ equivalent 
electrons in any symmetry $G$. Extensive use is made of symmetry adaptation 
techniques for the chain $O(3)^* \supset G^*$ (see [23]), 
where $G^*$ is the double group of the site symmetry group $G$. 

The starting point of this work relies on the transition 
moment for an $n$-photon absorption between 
an initial state $i$ and a final state $f$. In the framework of 
the electric dipole approximation, it is given by the 
well-known formula 
$$
M_{i \to f} = {1 \over \hbar^{n-1}} \sum_{ \left\{ v_j \right\} }
{ \left(f       \vert ({\cal E}_1 . \, D) \vert v_1 \right) 
  \left(v_1     \vert ({\cal E}_2 . \, D) \vert v_2 \right)
  \cdots 
  \left(v_{n-1} \vert ({\cal E}_n . \, D) \vert i   \right) 
\over 
(\Omega_i - \Omega_{v_1    }) 
(\Omega_i - \Omega_{v_2    }) \cdots
(\Omega_i - \Omega_{v_{n-1}}) } + {\rm permutations} 
\eqno (1)
$$
where the sum has to be extended over the intermediate 
states $v_j$ ($j = 1, 2, \cdots, n-1$) and {\it permutations} 
indicate that other terms must be added in order to take into 
account the $n$! permutations on the $n$ photons when they are 
different. In equation (1), the operator 
$({\cal E}_k. \, D)$ stands for the scalar product of the 
polarization vector ${\cal E}_k$ of the $k$-th photon 
($k = 1, 2, \cdots, n$) 
with the dipole moment operator $D$
for the $N$ electrons. Furthermore, 
the energy denominators have their usual meaning. 

At this stage, we have to make a sharp distinction between 
parity-allowed and parity-forbidden $n$-photon transitions. We shall 
examine in turn (in sections 2 and 3) these two types of 
transitions. In section 4, we shall report on intensity formulas for 
parity-allowed and parity-forbidden transitions. 

\noindent 
{\bf 2. Parity-allowed $n$-photon transitions}  

In this case, the parity is a good quantum number for the 
initial, final and intermediate states. In other 
words, the initial and final states have the same parity for 
$n$ even and opposite parities for $n$ odd. Among the 
parity-allowed $n$-photon transitions, we thus may have 
intra-configurational 
transitions for $n$ even and  
inter-configurational 
transitions for $n$ odd. Two 
particular cases are of special interest, viz., the 
intra-configurational two-photon transitions ($n=2$) and the 
inter-configurational one-photon transitions ($n=1$).

For $n \ne 1$, the sum over $\{v_j\}$ in (1) is very 
difficult to handle since the energy denominators depend on the 
internal structure of the intermediate states. This situation 
is analogous to that encountered in the study of one-photon 
(parity-forbidden) intra-configurational transitions. 
Therefore, we shall use here a 
quasi-closure approximation of the type introduced by Judd 
[1] and Ofelt [2] for $4f^N \to 4f^N$ one-photon transitions. As a net 
result, by using recoupling techniques in conjunction with a 
quasi-closure approximation, the transition matrix element (1) 
is amenable to the form 
$$
M_{i \to f} = ( f \Gamma' \gamma' \vert H_{\rm eff} \vert 
                i \Gamma  \gamma )   
\eqno (2)
$$
with 
$$
H_{\rm eff} = 
\sum_{\lambda_1, \lambda_2, \cdots, \lambda_{n-1}} 
    C[\lambda_1, \lambda_2, \cdots, \lambda_{n-1}] \, 
\left( 
\{ \cdots \{ \{ {\cal E}_1 {\cal E}_2 \}^{\lambda_1} 
                           {\cal E}_3 \}^{\lambda_2}
                    \cdots {\cal E}_n \}^{\lambda_{n-1}}
                                . \, U^{\lambda_{n-1}} \right)
\eqno (3)
$$
where $(.)$ indicates a scalar product. 
In equation (2), the 
labels $\Gamma$ and $\Gamma'$ stand for two irreducible 
representations of the group $G^*$. Furthermore, $\gamma$ and 
$\gamma'$ are multiplicity labels. The labels 
$\Gamma \gamma$ and $\Gamma' \gamma'$
are the only good quantum numbers for the states $i$ and $f$, 
respectively. 
In equation (3), the 
electronic part is contained in the Racah unit tensor 
$U^{\lambda_{n-1}}$ while the polarization 
dependence is contained in the tensor 
$\{ \ \}^{\lambda_{n-1}}$ 
(which describes the coupling of the $n$ polarization vectors 
associated with the $n$ photons). It should be noted that $\lambda_{n-1}$ 
cannot be equal to 0 (except for Rayleigh scattering). The 
$C$ parameters depend on 
the initial, final and intermediate configurations as well as the 
energies of the $n$ photons. 

In the limiting case $n=1$, equation (3) yields the operator 
$$
H_{\rm eff} = ( {\cal E} . \, D )
\eqno (4)
$$
of relevance for inter-configurational 
one-photon transitions. In the special case $n=2$, 
equation (3) returns the effective operator 
$$
H_{\rm eff} =            \sum_{\lambda = 1,2} 
                            C [\lambda] \, 
( \{ {\cal E}_1 {\cal E}_2 \}^{\lambda} . \, U^{\lambda} )
\eqno (5)
$$
that describes intra-configurational two-photon 
transitions in the Axe model [3]. 

\noindent 
{\bf 3. Parity-forbidden $n$-photon transitions} 

The $n$-photon transitions are forbidden 
(i.e., $M_{i \to f} = 0$) 
between states of the 
same parity for $n$ odd and between states of opposite parities 
for $n$ even. However, such transitions are observed 
in some cases. As an example, we have the case, treated 
independently by Judd [1] and Ofelt [2], of 
$4f^N \to 4f^N$ (intra-configurational) one-photon ($n=1$) transitions. 
Indeed, such transitions become weakly allowed owing to a violation 
mechanism induced by the (static or dynamic) crystal-field 
interaction. 

Going back to the case when $n$ is arbitrary, two parity violation 
mechanisms are possible in order to pass from $M_{i \to f} = 0$ 
to $M_{i \to f} \ne 0$~: 

(i) To use first-order time-independent perturbation theory in 
order to replace the initial, final and intermediate state 
vectors in (1) by state vectors with a not well-defined parity, 
the parity mixing being due to the odd crystal-field interaction. 

(ii) To start from the transition matrix element for a 
parity-allowed $(n+1)$-photon transition and to replace one of
the electric dipole operators (${\cal E}_k . \, D$) by the odd 
crystal-field interaction with appropriate permutations. 
%

We shall refer the mechanisms described by (i) and (ii) 
               to as $M[n,  1]$ and 
                     $M[n+1,0]$, respectively. The latter notation 
is a reminder 
       that $M[n,  1]$, respectively 
            $M[n+1,0]$, is concerned with 
$ n   ^{\rm th}$-order, respectively 
$(n+1)^{\rm th}$-order, time-dependent perturbation theory and 
$1$$^{\rm st}$-order,   respectively 
$0$$^{\rm th}$-order,   time-independent 
perturbation theory. It can be shown that, under some 
approximation, we have 
$M[n+1,0] \approx M[n,1]$ 
(see [21] for a detailed proof in the case $n=2$). 
 
By using either the $M[n,1]$ or $M[n+1,0]$ mechanism together with 
closure approximations and recoupling techniques, we can prove 
that the effective operator to be placed between the 
initial and the final ($0$$^{\rm th}$-order) state vectors 
in equation (2) reads 
$$
H_{\rm eff} = 
\sum_{\lambda_1, \cdots, \lambda_{n  }} \sum_{k} 
    C[\lambda_1, \cdots, \lambda_{n-1} ; k ; \lambda_n] \, 
    \left( \{ 
\{ \cdots \{ \{ {\cal E}_1 {\cal E}_2 \}^{\lambda_1} 
                           {\cal E}_3 \}^{\lambda_2}
                    \cdots {\cal E}_n \}^{\lambda_{n-1}} O^k \}^{\lambda_n}
                                   . \, U^{\lambda_{n}} \right)
\eqno (6)
$$
Equation (6) should be compared with (3). In equation (6), 
the electronic part is described by the Racah unit tensor 
$U^{\lambda_n}$. The polarization information is 
contained in the coupled tensor $\{ \ \}^{\lambda_n}$ where 
$O^k$ is a tensor whose components are defined from the odd 
crystal-field parameters [14]. 
The index $\lambda_{n-1}$ can take here the value 0, 
in contradistinction to parity-allowed $n$-photon transitions. 
The $C$ parameters here parallel the ones occurring 
in (3) but are given by a different formula. 

In the particular case $n=1$, equation (6) reads 
$$
H_{\rm eff} = 
\sum_{k         \, {\rm odd }} \, 
\sum_{\lambda   \, {\rm even}} \, 
                C[k ; \lambda] \, 
\left( \{ {\cal E} O^k \}^{\lambda} . \, U^{\lambda} \right) 
\eqno (7)
$$
which describes 
intra-configurational one-photon transitions, cf.~[1,2,14]. 
For $n=2$, equation (6) reduces to the effective operator
$$
H_{\rm eff} = 
\sum_{\lambda_1 = 0, 1, 2} \, \sum_{k \, {\rm odd}} \, 
\sum_{\lambda_2          } \, 
    C[\lambda_1 ; k ; \lambda_2] \, 
\left( \{ \{ {\cal E}_1 {\cal E}_2 \}^{\lambda_1} O^k 
\}^{\lambda_2} . \, U^{\lambda_2} \right) 
\eqno (8) 
$$
describing inter-configurational two-photon transitions 
[10,12,15,21]. 

\noindent 
{\bf 4. Intensity formula}

We are now in a position to give a formula for the intensity
$$
S_{  \Gamma  \to   \Gamma' } \; = \; \sum_{\gamma \gamma'} \; 
\left\vert M_{i(\Gamma \gamma) \to f(\Gamma' \gamma')} \right\vert ^2 
\eqno (9)
$$
of a (parity-allowed or parity-forbiddden) $n$-photon 
transition between the Stark levels $i$ of symmetry $\Gamma$ 
and $f$ of symmetry $\Gamma'$. In equation (9), the sums over 
$\gamma$ and $\gamma'$ have to be extended over all the 
components of the initial and final states, respectively. 

The calculation of $S_{  \Gamma  \to   \Gamma' }$ may be achieved, 
by making use of symmetry adaptation techniques [23], 
in the following way~: (i) express the initial and 
final states as well as the scalar products occurring in (3) 
and (6) in a form adapted to the chain of groups 
$O(3)^* \supset G^*$; (ii) apply the Wigner-Eckart theorem for the 
chain $O(3)^* \supset G^*$; (iii) use the factorization lemma 
for the coupling coefficients of the group $O(3)^*$ in an 
$O(3)^* \supset G^*$ basis; and, finally (iv) use the so-called 
orthogonality-completeness property for the Clebsch-Gordan 
coefficients of the group $G$. We thus obtain the intensity formula
$$
S_{\Gamma \to \Gamma'} = \sum_{ \left\{ k   _i \right\} } 
                         \sum_{ \left\{ \ell_i \right\} } 
                         \sum_{r} 
                         \sum_{s} \sum_{\Gamma''} 
I[ \left\{ k   _i \right\} 
   \left\{ \ell_i \right\}
r s \Gamma'' ; \Gamma \Gamma'] \; 
\sum_{\gamma''} \; 
P^{k   _{n-1}}     _{r        \Gamma'' \gamma''} \; \left( 
P^{\ell_{n-1}}     _{s        \Gamma'' \gamma''}    \right)^* 
\eqno (10)
$$
(with $1 \le i \le n-1$), 
the form of which holds for both parity-allowed and parity-forbidden 
$n$-photon transitions. For a derivation of (10), the reader may consult 
[18] in the case of intra-configurational 
parity-allowed   two-photon transitions and 
[21] in the case of inter-configurational 
parity-forbidden two-photon transitions.

The polarization dependence is clearly 
exhibited in (10) by the factors of type 
$$
 P^{\lambda_{n-1}} = \{ \cdots \{ \{ {\cal E}_1 {\cal E}_2 
                      \}^{\lambda_1} {\cal E}_3
                      \}^{\lambda_2} 
                              \cdots {\cal E}_n \}^{\lambda_{n-1}} 
\eqno (11)
$$
which already occur in (3) and (6). 
Of course, the intensity parameters $I$ are given by an 
expression that is specific of the kind of transition 
(parity-allowed or parity-forbidden) under consideration. The 
$I$ parameters depend on the initial and final states. They 
also depend on various reduced matrix elements 
$( \ \Vert \ \Vert \ )$ of spherical tensors and on several 
energy parameters (for the involved configurations and the $n$ 
photons). In the case of parity-forbidden transitions, 
they depend on the odd crystal-field parameters. 

The $I$ parameters present two interesting properties~: (i) an 
hermitian conjugation property and (ii) a factorization 
property when the symmetry group $G$ is multiplicity-free [22]. 
The number of $I$ parameters in the intensity formula (10) is 
partially controlled by property (i) and the following 
group-theoretical selection rules~: (i) the irreducible 
representation $\Gamma''$ of $G$ should be contained in the 
direct product $\Gamma'^* \otimes \Gamma$~ and (ii) the 
irreducible representations ($k_{n-1}$) and ($\ell_{n-1}$) of 
$O(3)$ should contain $\Gamma''$. Let us also emphasize that 
the number of $I$ parameters depends as well on the number of 
absorbed photons and that the higher the symmetry, the lower 
the number of $I$ parameters. 

\noindent 
{\bf 5. Discussion} 

In this paper, we have concentrated on $n$-photon absorption 
between Stark levels, with well-defined symmetry species, 
for a transition ion (as, e.g., a rare earth 
ion) in finite symmetry. We have obtained intensity formulas, 
for both parity-allowed and parity-forbidden $n$-photon 
transitions, which 
incorporate all the information arising from symmetry 
considerations. In the particular case $n=2$, our results 
constitute a quantitative counterpart to the qualitative 
treatment in [4,5]. (By quantitative we mean that 
the intensity parameters are given by expressions arising from 
well-defined mechanisms and are calculable from first principles.) 
The latter point shows that the tables 
of Bader and Gold [5] can be used for inter-configurational two-photon 
transitions as well as for intra-configurational two-photon transitions.  

The parity violation mechanism used in the present paper 
for parity-forbidden $n$-photon transitions is the same as the 
one introduced in the theory of Judd [1] and Ofelt [2] for 
intra-configurational ($4f^N \to 4f^N$) one-photon transitions. 
Our treatment of parity-forbidden $n$-photon transitions is an 
extension (to $n$ arbitrary), in a symmetry adapted form, of 
the one by Judd and Ofelt. 

For parity-allowed $n$-photon transitions, we have obtained 
results which extend, in a symmetry adapted form, 
the standard model of Axe [3] for intra-configurational 
two-photon transitions. 

We know that for intra-configurational two-photon 
transitions, 
some additional mechanisms (additional with respect to the 
second-order mechanisms arising from second-order 
time-dependent perturbation theory) have been introduced by 
various authors [6,7,9,11] in order to produce more 
efficient third- and fourth-order mechanisms. For instance, the 
third-order mechanisms introduced by Judd and Pooler [6] arise 
from second-order time-dependent perturbation theory plus 
first-order time-independent perturbation theory. Some 
additional mechanisms (including ligand-polarization effects) 
also may be introduced in the general 
case of parity-allowed and parity-forbidden 
$n$-photon transitions. This may lead to the replacement in 
equations (3) and (6) of the Racah unit tensor $U$ by some more 
complicated tensor with, for example, a spin- and orbit-dependence. 
The resulting intensity formula then assumes the 
same form as (10) (see [22]). 

The intensity parameters $I$ in (10) (involving possibly 
additional mechanisms) can be calculated in principle in an 
{\it ab initio} way. This yields, however, a very 
intricate quantum chemistry problem which requires the 
knowledge of precise wave-functions. Therefore, it is often 
interesting to consider them, or part of them, as 
phenomenological parameters, at least in a first approach. In 
this connection, it should be emphasized that a phenomenological 
approach may incoporate various mechanisms in a global way. 
This may be an advantage, as well as an inconvenience because 
it is not always feasible to distinguish the contributions 
arising from various mechanisms.  

Up to now, most of the experimental results for $n$-photon 
absorption are concerned with $n=1$ and $2$. The formalism used in the 
present paper has been applied for $n=2$ to various rare earth 
and transition-metal ions in crystals (see [13,16,17,19]). We hope that 
experimental data for polarization dependence of $n$-photon 
transitions will become available in order to test the model 
developed in this paper. 

\noindent 
{\bf Acknowledgements}

The authors are grateful to Dr.~G.W.~Burdick 
for a critical reading of the manuscript.

\noindent 
{\bf References}
\baselineskip 0.5 true cm 

\item{[1]} B.R. Judd, Phys. Rev.,      {\bf 127} (1962) 750.

\item{[2]} G.S. Ofelt, J. Chem. Phys., {\bf 37}  (1962) 511. 

\item{[3]} J.D. Axe, Jr., Phys. Rev.,  {\bf 136} (1964) A42.

\item{[4]} M. Inoue and Y. Toyozawa, J. Phys. Soc. Japan, 
{\bf 20} (1965) 363. 

\item{[5]} T.R. Bader and A. Gold, Phys. Rev.,         {\bf 171} (1968) 997.

\item{[6]} B.R. Judd and D.R. Pooler, J. Phys. C,      {\bf 15}  (1982) 591.

\item{[7]} M.C. Downer and A. Bivas,     Phys. Rev. B, {\bf 28}  (1983) 3677. 

\item{[8]} S.K. Gayen and D.S. Hamilton, Phys. Rev. B, {\bf 28}  (1983) 3706. 

\item{[9]} M.F. Reid and F.S. Richardson, Phys.~Rev.~B, {\bf 29} (1984) 2830. 

\item{[10]} R.C. Leavitt, Phys. Rev. B, {\bf 35} (1987) 9271. 

\item{[11]} L. Smentek-Mielczarek and B.A. Hess, Jr., 
Phys.~Rev.~B, {\bf 36} (1987) 1811. 

\item{[12]} A.G. Makhanek, V.S. Korolkov and L.A. Yuguryan, 
Phys. Status Solidi (b), {\bf 149} (1988) 231.

\item{[13]} J.C. G\^acon, J.F. Marcerou, M. Bouazaoui, B. Jacquier
and M. Kibler, Phys. Rev. B, {\bf 40} (1989) 2070. 

\item{[14]} M. Kibler and J.C. G\^acon, Croat. Chem. Acta, {\bf 62}  (1989) 783.

\item{[15]} J. Sztucki and W. Str\c ek, Chem. Phys.,       {\bf 143} (1990) 347.

\item{[16]} J.C. G\^acon, B. Jacquier, 
J.F. Marcerou, M. Bouazaoui and M. Kibler, J. Lumin., {\bf 45} (1990) 
162. 

\item{[17]} J.C. G\^acon, M. Bouazaoui, B. Jacquier, M. Kibler, L.A. 
Boatner and M.M. Abraham, Eur. J. Solid State Inorg. Chem., 
{\bf 28} (1991) 113. 

\item{[18]} M.R. Kibler, in W. Florek, T. Lulek and M. Mucha 
(eds.), Symmetry and Structural Properties 
of Condensed Matter, World Scientific, Singapore, 1991, p. 237. 

\item{[19]} J. Sztucki, M. Daoud and M. Kibler, Phys. Rev. 
B, {\bf 45} (1992) 2023. 

\item{[20]} G.W. Burdick and M.C. Downer, 
in A.H.~Kitai (ed.), Visible Luminescence, Chapman and 
Hall, London, in the press.

\item{[21]} M. Daoud and M. Kibler, Proc. International Workshop on 
Laser Physics, Dubna, Russia, April 1992, submitted for publication. 

\item{[22]} M. Daoud, Doctorate Thesis, in preparation. 

\item{[23]} M.R. Kibler, in J.C.~Donini (ed.), Recent Advances in Group 
Theory and Their Application to Spectroscopy, 
Plenum Press, New York, 1979, p.1. 

\bye